\documentclass[11pt]{article}

\usepackage[utf8]{inputenc}
\usepackage[margin=2cm]{geometry}
\usepackage{graphicx}
\usepackage{setspace}

\usepackage{amsmath}
\usepackage{amssymb}
\usepackage{url}
\usepackage{hyperref}
\usepackage{xcolor}
\usepackage{tikz}
\usepackage{xcolor}
\usepackage{tabularx}
\usepackage{titlesec}
\usepackage[defaultlines=2,all]{nowidow}

\titlespacing{\paragraph}{0pt}{0pt}{1ex}

\usepackage[margin=2cm,font=footnotesize]{caption}

\usetikzlibrary{arrows,shapes}
\tikzstyle{vertex}=[circle,fill=black!10,minimum size=20pt,inner sep=0pt]
\tikzstyle{ego} = [vertex, fill=black!50]
\tikzstyle{edge} = [draw,thick,-]

\usepackage[backend=biber,citestyle=authoryear,isbn=false,url=false,eprint=false,sorting=none,giveninits=true,maxcitenames=1,uniquename=init]{biblatex}
\AtEveryBibitem{
  \clearfield{note}
}

\definecolor{bleucite}{RGB}{34,111,212}
\usepackage{hyperref}
\hypersetup{hidelinks}
\AtEveryCite{\color{bleucite}}

\title{The anatomy of a population-scale social network}
\author{Eszter Bokányi$^{1,*}$, Eelke M. Heemskerk$^1$, Frank W. Takes$^2$}

\date{{\footnotesize
    $^1$University of Amsterdam\\
    $^2$Leiden University\\
    $^*$\href{mailto:be.bokanyi@uva.nl}{e.bokanyi@uva.nl}}}

\begin{document}

\onehalfspacing

\maketitle

\abstract{
Large-scale human social network structure is typically inferred from digital trace samples of online social media platforms or mobile communication data. 
Instead, here we investigate the social network structure of a complete population, where people are connected by high-quality links sourced from administrative registers of family, household, work, school, and next-door neighbors. 
We examine this multilayer social opportunity structure through three common concepts in network analysis: degree, closure, and distance. 
Findings present how particular network layers contribute to presumably universal scale-free and small-world properties of networks. 
Furthermore, we suggest a novel measure of excess closure and apply this in a life-course perspective to show how the social opportunity structure of individuals varies along age, socio-economic status, and education level. 
Our work provides new entry points to understand individual socio-economic failure and success as well as persistent societal problems of inequality and segregation. 
}
\setlength{\parindent}{0em}
\setlength{\parskip}{0.8em}

\section{Introduction}

Recent studies on large-scale human social networks successfully utilize the growing abundance of digital trace data such as online social media or mobile communication datasets to uncover fundamental insights on human interactions \parencite{bailey2018social,asikainen2020cumulative,chetty2022social1,chetty2022social2}. 
However, deriving scientific meaning from large-scale data originally not created for the purpose of academic research is far from trivial. 
In networks created from such data, nodes are a mere sample based on users of a certain platform or mobile phone provider.  
Therefore, they are likely not representative of investigated populations.
Moreover, most social ties ultimately appear as indistinguishable edges after being inferred from messages  \parencite{leskovec2008planetaryscale}, calls \parencite{onnela2007analysis,blondel2015survey}, or friendship nominations \parencite{corten2012composition, chetty2022social1, chetty2022social2}.
This inference is often based on non-generalizable research design choices \parencite{peel2022statistical}, for example, when mobile calls are thresholded by call volume or call numbers to arrive at a set of social ties. 
The formation of edges from online social networks might also be influenced by platform-specific recommendation algorithms. On most platforms, edges represent people's activity rather than their social connectivity, or may not cover important connections such as close family.
As a result, most large-scale social network datasets are non-random samples of an underlying social structure \parencite{leskovec2008planetaryscale, mislove2011understanding, corten2012composition}.  
It is therefore hard to derive generically applicable actionable insights from such data, or to generalize findings across studies \parencite{lazer2021meaningful}.
Despite the above mentioned non-trivial concerns, online and mobile communication social network data are still used for instance as a direct proxy for real-world friendships, or to inform policy measures on crucial issues as combating prevalent societal problems such as inequality \parencite{toth2021inequality, chetty2022social1}.

In this paper, we present a promising way forward in obtaining reliable insights into large-scale social structures through a population-scale social network analysis based on government-curated administrative register data.
While such a network is based on national registers that are initially also not designed for academic research, its coverage of an entire population, explicit notion of well-categorized edge types, and the presence of high-quality demographic information provide insights at an unprecedented level of detail.
Most importantly, nodes are people officially enlisted in a country's administrative system. 
This gives a clear geographical boundary and legal definition of the included nodes resulting in an almost perfect population-scale coverage and thus, a representative node sample.
As opposed to more informal ties such as retweets, phone calls, or friendships that are typically captured in aforementioned online social media data, social ties encoded in official registers are \emph{formal ties}. However, it has been shown that formal ties do capture the majority of people's strong connections \parencite{vaneijk2010unequal,wrzus2013social, buijs2022friends}.
Using formal ties, we can describe individuals' \emph{social opportunity structure} similarly to, e.g., labour opportunities \parencite{tickamyer1990poverty}.

The register-based data allows us to empirically take into account how different edge types contribute to social opportunity structures.
The framework used in this work is hence a node-aligned multilayer social network \parencite{kivela2014multilayer, murase2014multilayer, boccaletti2014structure}, in which each edge type is modeled as a separate network layer. Previous work pointed towards the necessity of distinguishing edge types in order to adequately understand overlapping community structure of social networks \parencite{ahn2010link}, or the measurement and modeling of weak ties and their importance in network cohesion \parencite{onnela2007structure, murase2014multilayer}.
For automatically collected large-scale networks, it is sometimes possible to infer edge types from the communication content \parencite{deri2018coloring} or to capture different interactions between nodes by the data collection design \parencite{szell2010multirelational,jankowski2017multilayer}.
However, these examples remain exceptions and limited to the particular measurement setting used in these studies,
or include only a small number of nodes (see \cite{socievole2014wireless}, \cite{bahulkar2018interaction}, and \cite{dickison2016multilayer}, Chapter 3).
Instead, we map and analyse the multilayer network properties of an entire country's population.
An important benefit of register data is that it provides a wealth of reliable demographic information on the nodes that is often unavailable when other data sources are used, for instance on income, education, age, or neighbourhood properties.
Therefore, not only does this approach deliver an efficient way for mapping people's social opportunity structures at a high level of granularity, but it also holds the promise of  deriving meaningful conclusions about consequences of social structure for crucial matters.
These include quantifying poverty and inequality \parencite{dimaggio2012network, toth2021inequality}, constructing better epidemic models \parencite{buckee2021thinking}, understanding related public health issues \parencite{smith2008social, cullati2018development}, and growing political polarization \parencite{conover2011political}.

We present a network analysis of all 17.2M residents registered in the Netherlands.  The network is sourced from Statistics Netherlands and contains roughly 827M edges of the following types:  close and extended family, household, next-door neighbor, school, and work relationships. We present a thorough investigation of the anatomy of this network, and find three surprising differences with respect to commonly observed or expected properties of other well-known large-scale social networks.
First, the fat-tail degree distribution of online social networks \parencite{barabasi1999emergence, mislove2007measurement} is conspicuously absent because the generative logic of connectivity and therewith the degree distribution is different in a population-scale social network of formal ties.
Second, short average shortest path length is generated not through random link rewiring \parencite{watts1998collective} or strong edges closing large network distances \parencite{park2018strength}, but through a particular interaction of the various network layers. 
Third, closure measured by the local clustering coefficient \parencite{ahn2007analysis} reaches unprecedented high levels, and increases with increasing degree.

To address the challenges in the measurement of closure, we propose a novel multilayer measure that we call \emph{excess closure}, which specifically captures social closure that is the result of overlap between edges in different layers. 
An application of this measure on the population-scale social network reveals fascinating patterns of social opportunity structures in the Netherlands. 
We show that across different age groups, higher degree characterizes higher income and higher education population subgroups, and that degree peaks for people with a university age. 
Similarly, excess closure is high for the youngest and the oldest, but reaches its lowest point for young adults. 
Interestingly, this behaviour is markedly different for different socio-economic groups. 
Altogether, our results illustrate the need for new measurement tools tailored to the particular multilayer structure of population-scale social networks.
Moreover, our work brings the field a step closer to reliably using large-scale social network data for deriving policy-relevant actionable insights from the network structure underlying our complex society.

\section{Results}
\label{sec:results}

The population-scale network data consists of nodes that correspond to the roughly 17.2M people (the whole Dutch population) registered in the Netherlands on October 1, 2018 \parencite{vanderlaan2022person}, as well as six sets of edges derived from various government registers, referred to as network layers. 
In particular, we consider Close Family $C$ and Extended Family $E$ derived from the parent-child and partner registers, and Household $H$, Work $W$, and School $S$,  being projections of people's affiliations.
To avoid confusion between next-door neighbor relationships and network neighbors, i.e., two adjacent nodes connected by an edge, we refer to edges from the layer of people's connectivity based on spatially close households as Next-door, abbreviated by $N$. 
As such, our network is a population-scale node-aligned multilayer network with a single aspect \parencite{kivela2014multilayer}, where each layer contains the same set of nodes, and edges only connect nodes within the layers.
All data was pseudonymized, and analysis took place in a GDPR-compliant manner, ensuring that the identity of individuals is not revealed (see \nameref{sec:methods} section and the Ethical statement for details).

We start with an exploration of the network topology, revisiting three well-known structural features of social networks: degree, clustering, and shortest path length. 
The purpose is to establish to what extent the topological features of this population-scale social network resemble those of other previously studied large-scale social networks, with a special interest in how the different network layers contribute to these measures.
We then use these insights to characterize the social opportunity structures of different demographic groups by age, income, education level, and urbanization level of their home locations.

\subsection{Degree}

Degree distributions of large-scale social networks from the literature typically are said to exhibit two main features.
First, most nodes have a relatively low degree (number of connections), meaning that the lowest degrees are the most abundant in these networks. Second, some nodes have extremely large degrees, i.e., there is a fat-tailed degree distribution \parencite{corten2012composition, leskovec2008planetaryscale, mislove2007measurement, myers2014information}.
In this section, we revisit whether this holds for our population-scale network of formal connections, and show how the degree distributions and generating mechanisms of different layers contribute to the overall shape of the total degree distribution.

\begin{figure}[!b]
    \centering
    \includegraphics[width=0.9\textwidth]{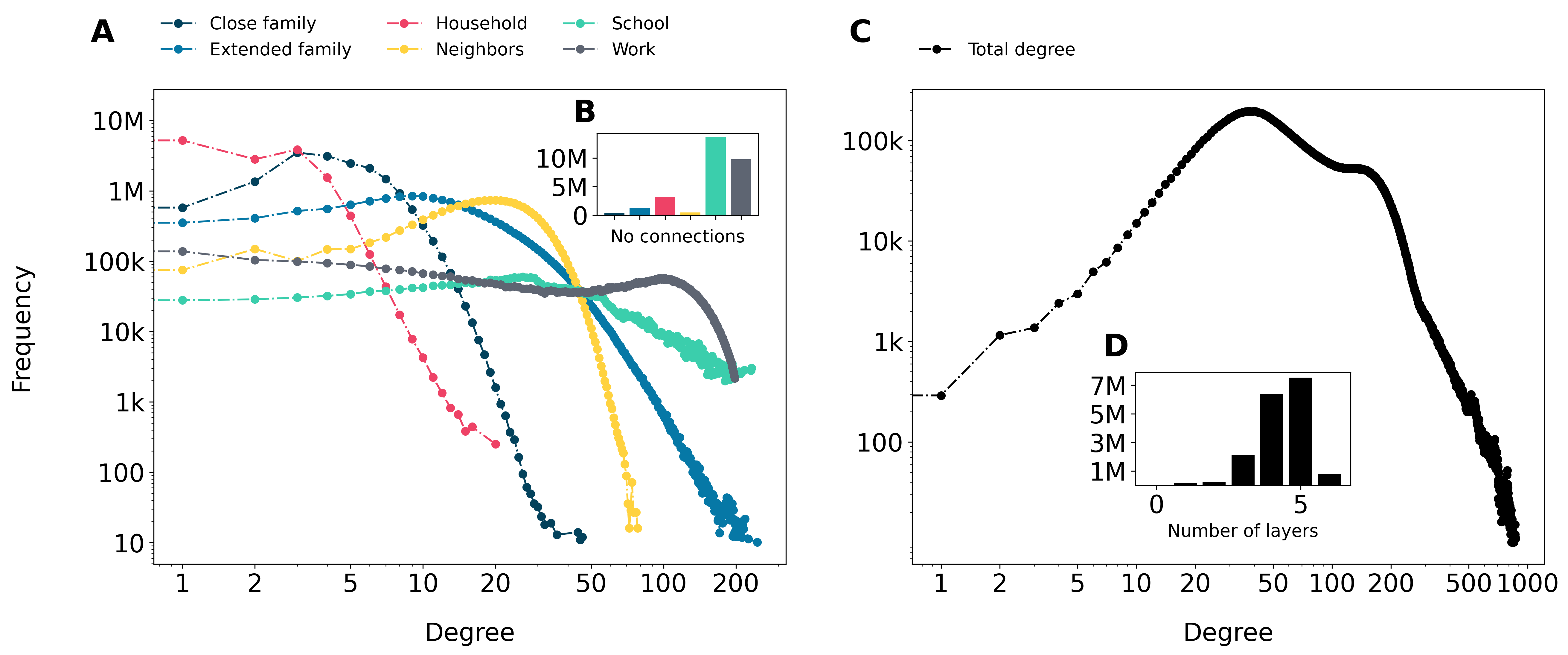}
    \caption{\textbf{Degree distributions per layer and in total.} (A) The degree distributions for the individual layers, distinguished by colors. 
    (B) Number of nodes with degree zero in each layer. (C) Distribution of the total degree, on a double logarithmic scale. (D) Distribution of the number of layers in which a node has nonzero degree. From all figures, categories containing less than 10 nodes are omitted for privacy reasons.
    }
    \label{fig:degree_distribution}
\end{figure}

Figure~\ref{fig:degree_distribution} shows the degree distributions for the six different layers, the total degree distribution, and the distribution of the number of unique layer types each node is involved in (see \nameref{sec:methods} section for details on how missing data was handled).
Figure~\ref{fig:degree_distribution}B, the left inset, shows the number of nodes with zero degree in these layers. Approximately 2.5\% of the nodes have no close family members, and 18.6\% live in a one-person household having a degree of 0 in the household layer.
The number of nodes with degree 0 in the extended family layer is almost three times as much, 7.6\% of all the nodes.
These nodes mostly correspond to first-generation migrants coming either alone, or only with their close family members to the country.
Apart from these isolated nodes, the degree distributions of the close family, extended family, and household layers for nodes with at least one neighbor in Figure~\ref{fig:degree_distribution}A all illustrate that a low number of connections is very common.
After this initial flat regime, all three distributions show a somewhat fat fail corresponding to a relatively small group of people belonging to quite large close or extended families, and that there are some large households.
These layer-wise distributions thus resemble fat-tailed distributions from communication networks or that of online social relationships.

The degree distribution of the school layer indicates typical class sizes in the primary and secondary education that peak around 20, and there is a slowly decaying tail towards the degrees of large university yeargroups. The number of colleagues, that corresponds to the number of neighbors in the work layer, starts off with a flat distribution in the lower regime, and then shows a peak slightly above 100 colleagues (the latter is because of the underlying data generation mechanism, see \nameref{sec:methods}).

Interestingly, when adding up the degrees of different layers for the nodes to obtain the total degree distribution shown in Figure~\ref{fig:degree_distribution}C, there are hardly any nodes (0.08\%) that remain disconnected, while the highest degree peaks at only 844. 
Between the two extremes, 80\% of the nodes  are in the range of 28 to 177 connections. 
The peak of the work layer at a degree of about 100 remains visible. 
Most nodes have edges in 4 to 6 layers, as shown in  Figure~\ref{fig:degree_distribution}D. 
Unlike many previously studied social networks, the total degree distribution follows none of the most commonly assumed shapes, i.e., it is neither a power-law  nor a lognormal distribution. 
This suggests that in a network where the existence of an edge does not involve user activity, as expected, preferential attachment is not a generating mechanism behind degrees.

\begin{table}[h!]
    \centering
    \begin{tabular}{lrrrrrr}
 {} & $C$ & $E$ & $H$ & $N$ & $S$ & $W$\\
$C$ & 1.00 & 0.52 & 0.10 & 0.03 & -0.14 & 0.08 \\
$E$ &   & 1.00 & 0.06 & 0.01 & -0.04 & 0.08 \\
$H$ &   &   & 1.00 & 0.07 & 0.10 & -0.01 \\
$N$ &   &   &   & 1.00 & 0.04 & 0.07 \\
$S$ &   &   &   &   & 1.00 & -0.05 \\
$W$ &   &   &   &   &   & 1.00 \\
    \end{tabular}
    \caption{\textbf{Inter-layer degree assortativity.} Pearson correlations between degrees calculated from different layers: close family ($C$), extended family ($E$), household ($H$), neighbors ($N$), school ($S$), and work ($W$). Only nodes with complete family information are included.}
    \label{tab:degree_corr}
\end{table}

To understand how the degree distributions of the several layers add up into the total degree distribution of Figure~\ref{fig:degree_distribution}C, we show in Table~\ref{tab:degree_corr} the weighted correlation between the degrees of people for whom the family information was complete.
As expected, the number of close family members and extended family members show a strong positive correlation of 0.53.
A somewhat weaker but still positive significant relationship is that the larger the close family, the greater household a person lives in, which is shown by a correlation of 0.10.
The fraction of working-age adults is also higher in larger families, therefore, the close family degree is positively correlated to work degree, and negatively to school degree.
The same effect is existent but weaker for the extended family degree. However, large households typically house more school-aged children or young people.
Therefore, despite opposite correlation signs of household size to family size and school degree, school degree and household degree are positively correlated. 
To sum  up, the different layers contribute in their own distinct manner to the total degree distribution, reflecting the different generating mechanisms in those layers.

Overall, these results show that merely observing total degree distributions as in Figure~\ref{fig:degree_distribution}C without taking into account the multilayer nature of large-scale social networks is disadvantageous. 
This is particularly the case because it poses the risk of attributing meaning to compounded distributions. 
Moreover, a deeper understanding of the generating mechanisms of edges and the correlation between degrees in different layers is necessary for the correct interpretation of patterns in the total degree distribution.

\subsection{Shortest path lengths}

We continue with looking at how the different layers play a role in realizing shortest paths. The seminal work of Watts and Strogatz suggests that, starting from a locally clustered ring-like structure, the addition of a limited number of randomly rewired edges creates a low diameter and low average shortest path lengths \parencite{watts1998collective}. 
For large social networks that cover a significant portion of a population, \textcite{park2018strength} argues that the small-world structure is created by so-called ``network wormhole" edges. These are relatively strong ties that connect previously distant parts of the network.
In what follows, we assess  whether the Dutch population-scale social network has small-world properties, and whether network wormholes are a driving force behind it.

To measure the effect of different layers on the average shortest path length, we create networks from different subsets of layers (or in some cases, a subset of edges within a layer, denoted by a subscript).
The created networks range in density from only taking parent-child directed edges from the close family layer $C_p$ to the full six-layer network.
We subsequently add layers to the parent-child edges $C_p$, being all close family links $C$, the extended family layer $E$, the household layer $H$, next-door neighbors $N$, primary school $S_p$ and secondary school $S_s$, the entire school layer $S$, small workplaces ($W_s$ where workplace size $\le$ 50), and at last, every edge from all six layers ($C+E+H+S+W+N$).
Table~\ref{tab:sp} summarizes the average shortest path lengths for these networks along with several other key global network measures.

\begin{table}[!b]
\centering
\begin{tabular}{lrrrrrr}
Layers & Nodes & Edges & Components & $GC$ & $D$ & $\overline{d}$ \\
$C_p$ & 15.76M & 19.16M & 963.75k & 0.58 & 288 & 75.03 \\
$C+E+H$ & 16.80M & 162.15M & 317.30k & 0.93 & 33 & 9.66 \\
$C+E+H+N$ & 17.25M & 361.95M & 2.77k & 1.00 & 64 & 6.12 \\
$C+E+H+N+S_p$ & 17.25M & 435.82M & 2.46k & 1.00 & 63 & 5.61 \\
$C+E+H+N+S_p+S_s$ & 17.25M & 478.66M & 2.32k & 1.00 & 62 & 5.39\\
$C+E+H+S+N+W_s$ & 17.25M & 496.34M & 1.83k & 1.00 & 42 & 5.25 \\
$C+E+H+S+N+W$ & 17.25M & 826.99M & 1.15k & 1.00 & 22 & 4.64 \\ 
\end{tabular}
\caption{\textbf{Network properties for various layer combinations.} 
Columns show the number of non-isolated nodes, number of edges,  number of connected components, relative size of the giant component $GC$, this component's diameter $D$ and estimated average shortest path length~$\overline{d}$.}
\label{tab:sp}
\end{table}

First, we observe that average distances in the parent-child layer are quite large: two randomly selected nodes are roughly 75 steps away from each other.
Adding close family and extended family alongside with household links to this ``backbone" of parent-child links already decreases the average distance to below 10.
This addition corresponds to shortcutting paths of length 2, 3, 4, or even larger in the parent-child network, to distance 1.
For example, a sister would have been at distance 2, and a cousin would have been at distance 4, but in the direct extended family layer they are directly connected, so at a distance of 1.
These family structures can be seen as long ``chains'' of highly clustered bubbles of relatives merged together by marriages, partnerships and having common children.
Adding next-door neighbor relationships on top of family and household edges shortens the average shortest path length by an additional 3.54 steps down to around 6.12.
Primary schools decrease this distance to 5.61, with small workplaces and other school types add subsequent decreases, leading to an average shortest path length of 5.25.
Thus, next-door neighbor, school and work relationships are effective in connecting more distant parts of the network, and deliver an average shortest path length that go below the characteristic value we would expect in so-called small-world network \parencite{milgram1967small,watts1998collective}. Including even more sporadic potential work connections, that is, all of the larger workplaces, further decreases average distances to 4.64, that approaches the value suggested for Facebook by \cite{backstrom2012four}. It also provides empirical confirmation of the findings of \parencite{dodds2003experimental}, who suggest that shortest paths
in large-scale networks are likely to be mediated by professional or school connections.

\begin{figure}[!t]
    \centering
    \includegraphics[width=0.9\textwidth]{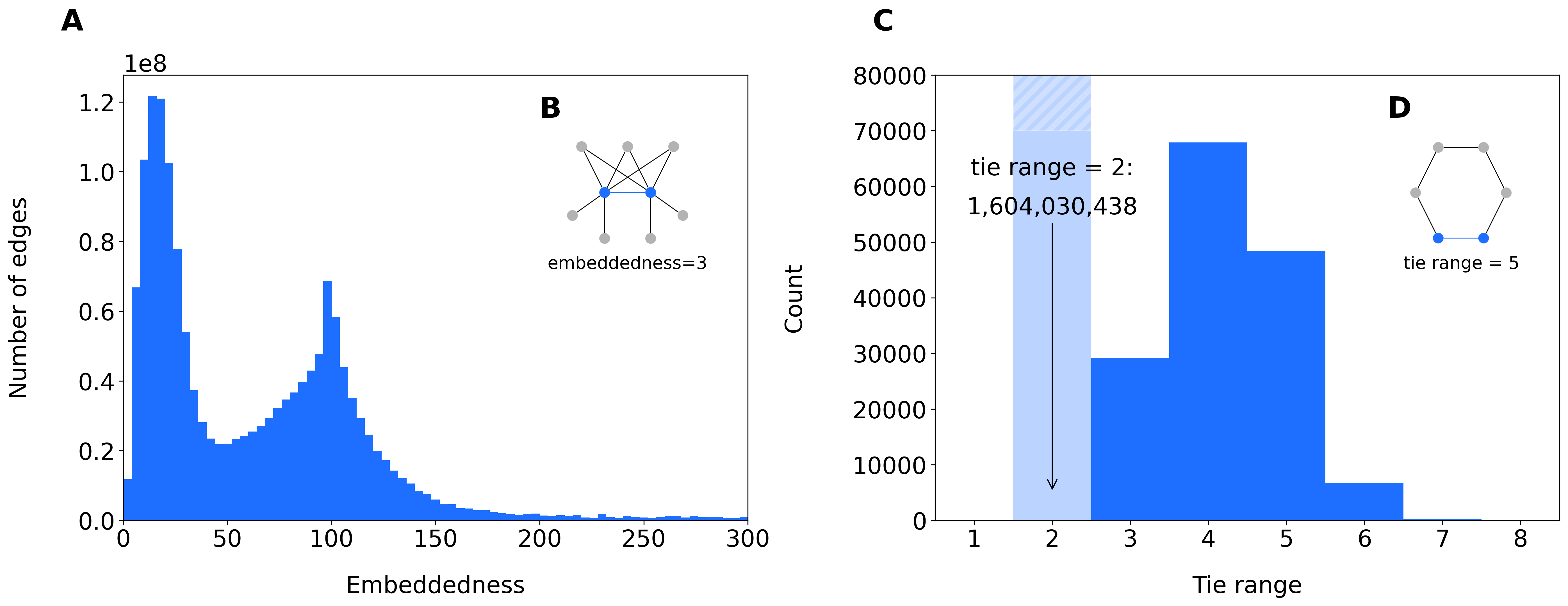}
    \caption{\textbf{Embeddedness and tie range.} (A) Embeddedness distribution of all edges in the network. (B) Example of an edge with embeddedness 3. (C) Tie range distribution for edges that are not part of a triangle, so with embeddedness 0. (D) Example of an edge with tie range~5.}
    \label{fig:embeddedness_tie_range}
\end{figure}

The question that remains is whether the underlying mechanism is through the aforementioned wormhole edges. This  can be investigated by the measures of edge embeddedness and tie range.
Edge embeddedness measures the number of overlapping neighbors of the pair of nodes that are connected by the edge (see Figure~\ref{fig:embeddedness_tie_range}B).
Local bridges that shorten average distances are edges for which the nodes at the two endpoints have no common neighbors, and thus correspond to an embeddedness of zero (see the \nameref{sec:methods} section for details).
Figure~\ref{fig:embeddedness_tie_range}A shows the distribution of embeddedness in our network, with the peak around 100 corresponding to the workplace sampling process (as explained in the \nameref{sec:methods} section). 
But most interestingly, there are only around 312k local bridge edges (0.02\%) in our network having zero embeddedness.

For these local bridges, we measure the length of the second shortest path between nodes corresponding to an edge (see Figure~\ref{fig:embeddedness_tie_range}D), the so-called tie range (see \nameref{sec:methods} section).
Note that this value is one less than the length of the shortest cycle going through the selected edge.
Therefore, edges with a nonzero embeddedness automatically have a tie range of 2.
Network wormholes are relatively strong long-range ties having a range of at least 6 \parencite{park2018strength}.
In the Twitter network of the country of Singapore, these ties amounted up to 0.46\% of all edges.
Figure~\ref{fig:embeddedness_tie_range}B shows the distribution of tie range in our network, that mostly spans small distances of either 2 or 3.
Only 9419 edges have a range of at least~6, making the proportion of such wormhole edges $0.0011\%$, two orders of magnitude less than in the case of aforementioned Twitter network. A closer look at the data shows that 92\% of these long-range edges are either close family, extended family, or household edges. 
This implies that such edges in themselves are not expected to create smaller shortest path lengths in the Dutch population-scale network.

Thus, we conclude that in our population-scale social network, small shortest path lengths are not brought about by wormhole edges, but instead appear to be facilitated by a particular multilayer interaction where next-door neighbor, school, and work edges cut paths of clustered kinship relationships shorter.

\subsection{Clustering and excess closure}

Finally, we turn to closure as an important property of the social opportunity structure of individuals.
Throughout people's lives and across demographic groups, closure  affects their access to opportunities and information~\parencite{borgatti2009network, charoenwong2020sociala, toth2021inequality}.
It is well documented that human social networks are more clustered than what could be expected from sensible null models \parencite{erdos1960evolution, barabasi1999emergence}.
Closure is typically measured as the average of the local clustering coefficient over all nodes, which captures the fraction of closed triangles between a node's direct neighbors. 
For Facebook and Twitter, both very large online social networks, the clustering coefficient has an average value of 0.4 and 0.23 for degree $k=5$, 0.30 and 0.19 for degree $k=20$, and around 0.14 for degree $k=100$ \parencite{ugander2011anatomy, myers2014information}.
In these and similar networks, the local clustering coefficient is said to decay roughly with $k^{-1}$, thus, clustering mostly disappears for large degrees, especially for large hubs \parencite{ahn2007analysis, leskovec2008planetaryscale, corten2012composition}.

\textbf{Local clustering coefficient}.

\begin{figure}[!t]
    \centering
    \includegraphics[width=0.9\textwidth]{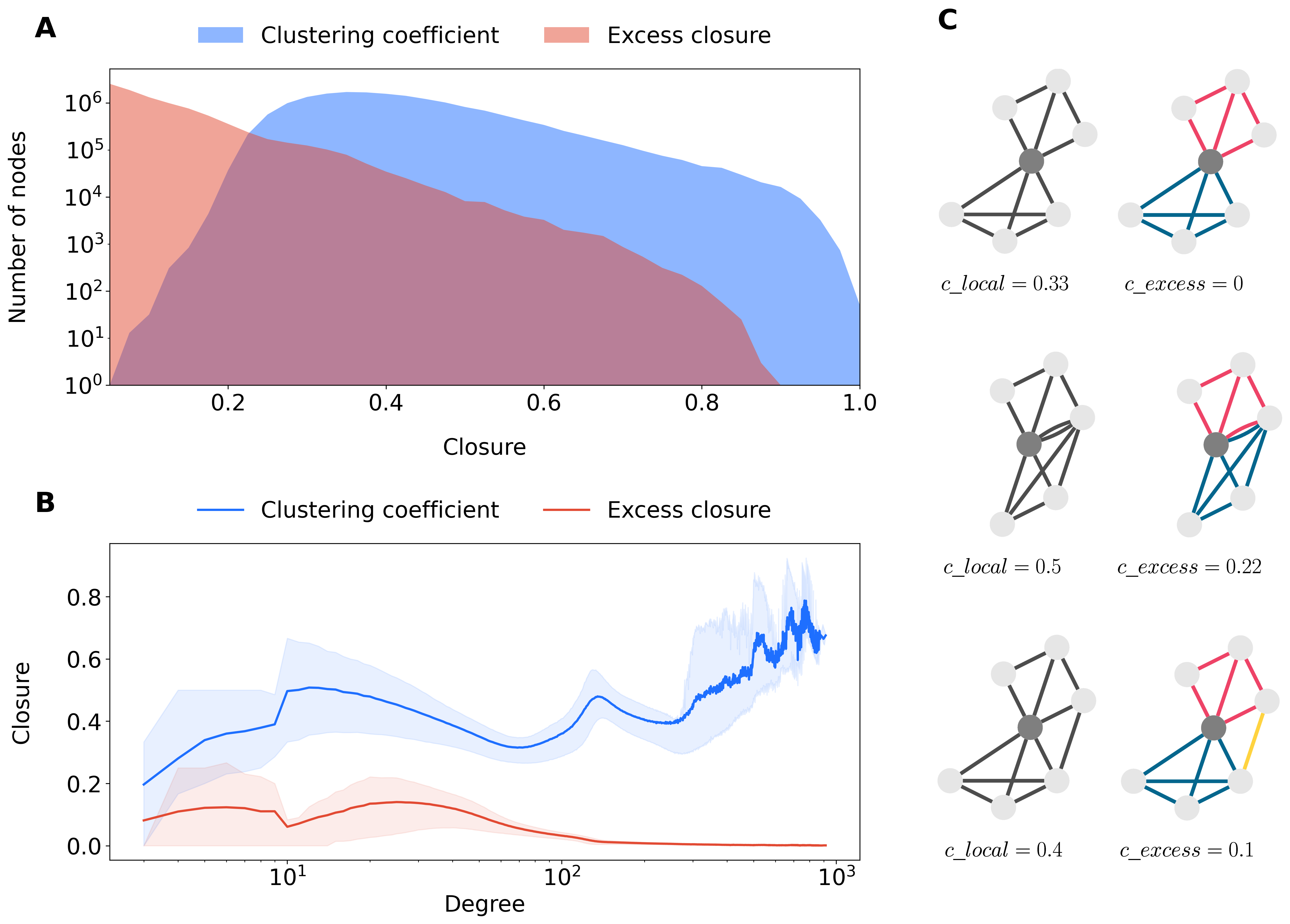}
    \caption{\textbf{Local clustering coefficient and excess closure.} (A) Distribution of the local clustering coefficient and excess closure.
    (B) Average local clustering coefficient and average excess closure for nodes with a certain degree. Shaded areas corresponds to the 25th and 75th percentiles within that degree. 
    (C) Local clustering coefficient of dark grey node, calculated on all network layers (left column, grey edges), and excess closure (right column, colored edges).}
    \label{fig:closure}
\end{figure}

Contrary to findings described in the literature discussed above, in our network, the local clustering coefficient \parencite{watts1998collective} reaches unprecedented high levels.
This is shown by the distribution in Figure~\ref{fig:closure}A in blue, where half of the nodes have a local clustering coefficient value $c\_{local}$ larger than 0.38, with an average of 0.40.
If measured as a function of degree (Figure~\ref{fig:closure}B), another peculiar observation in our network is that the local clustering coefficient actually increases (with the exception of only two apparent local minima) with growing degree.
This is caused by large degrees originating from layers that are in fact bipartite projections of particular affiliations (school, work), where the larger the degree, the more triangles are present in the ego network structures.
Furthermore, close and extended family are also highly clustered, yet they are so by nature, regardless of degree. The origin for this lies in the logic of kinship relations and how people maintain them \parencite{hamberger2011kinship,david-barrett2019network}.
High clustering also holds for next-door neighbors, because being spatially close is often a transitive relationship.

We conclude, based on the observations above, that the local clustering coefficient is not a useful measure for understanding closure in the considered population-scale social network. 
Moreover, it inherently disregards the multilayer topology of the network.
This is problematic because while the local clustering coefficient is expected to be high inside each layer, we currently do not capture the ``unexpected" closure that comes from the complex interaction of various layers in the network.

\textbf{Excess closure}. 
Figure~\ref{fig:closure}C shows three small typical examples of local structures in the left column, all exhibiting high levels of local clustering.
However, this degree of clustering merely reflects that the neighborhoods of nodes contain many closed triangles within the layers, that can be \emph{expected} because of the generative mechanisms behind the edge creation processes.
To discount for this abundance of \emph{expected} within-layer triangles, we propose a modification of the local clustering coefficient that we call the \emph{excess closure}, denoted by $c\_{excess}$. The excess closure measure $c\_{excess}$ is a value between 0 and~1, where  0 means that there is no overlap in the neighbors within different layers of an ego.
A value close to 1 would happen when the number of triangles with edges from at least two layers approaches the number of wedges (open triangles) that are not closed by same-layer edges.
It thus particularly measures closure brought about by what could be called multilayer triangles (see \nameref{sec:methods} for details). 

Excess closure can either exist because neighboring nodes have edges in multiple layers, for example when a next-door person is also a schoolmate, or because of triadic closure between edges from different layers.
The examples of the right hand column of Figure~\ref{fig:closure}C illustrate this by highlighting the layers in different colors.
In the first case, clustering only happens within layers, therefore, excess closure is 0.
In the second case, there is an alter in the ego network connected by two edges from two layers.
Thus, there are some triangles that contain edges from at least two layers, that add to the excess closure measure value. 
In the bottom example, there is a local bridge between the alters connected by edges from two highly clustered layers that realizes excess closure.

The distribution of excess closure values $c\_{excess}$ for all 17.2M people in the population-scale network is shown in red in Figure~\ref{fig:closure}A.
Compared to the local clustering coefficient, excess closure indeed has typically much lower values.
However, it certainly happens that network layers and thus social opportunity structures overlap or are bridged.
Figure~\ref{fig:closure}C illustrates that contrary to the local clustering coefficient, excess closure does not counterintuitively increase with degree.
All in all, excess closure overcomes the problem of very large local clustering coefficients that originates from the generating mechanisms of formal ties.
More specifically, it is suitable for uncovering closure caused by overlap and local bridging in social opportunity structures.

\subsection{Network positions in different stages of life}

A unique advantage of register-based population-scale social network data is that high quality census information is available on the individuals. This allows us to investigate how social network position of individuals differ across, for instance, income, school level, or living environment. 
In addition, the complete coverage of individuals of all ages allows us to take a life-course perspective, investigating how social network positions are different across age groups.
We present results of such an analysis as an illustration of our population-scale network's potential as a foundation for detailed computational social scientific modeling of processes such as persisting inequality~\parencite{toth2021inequality,chetty2022social1, chetty2022social2}, segregation~\parencite{vanderlaan2022whole} or various dynamical phenomena~\parencite{boccaletti2014structure} that may play out on the network structure. 

\begin{figure}[!b]
    \centering
    \includegraphics[width=0.9\textwidth]{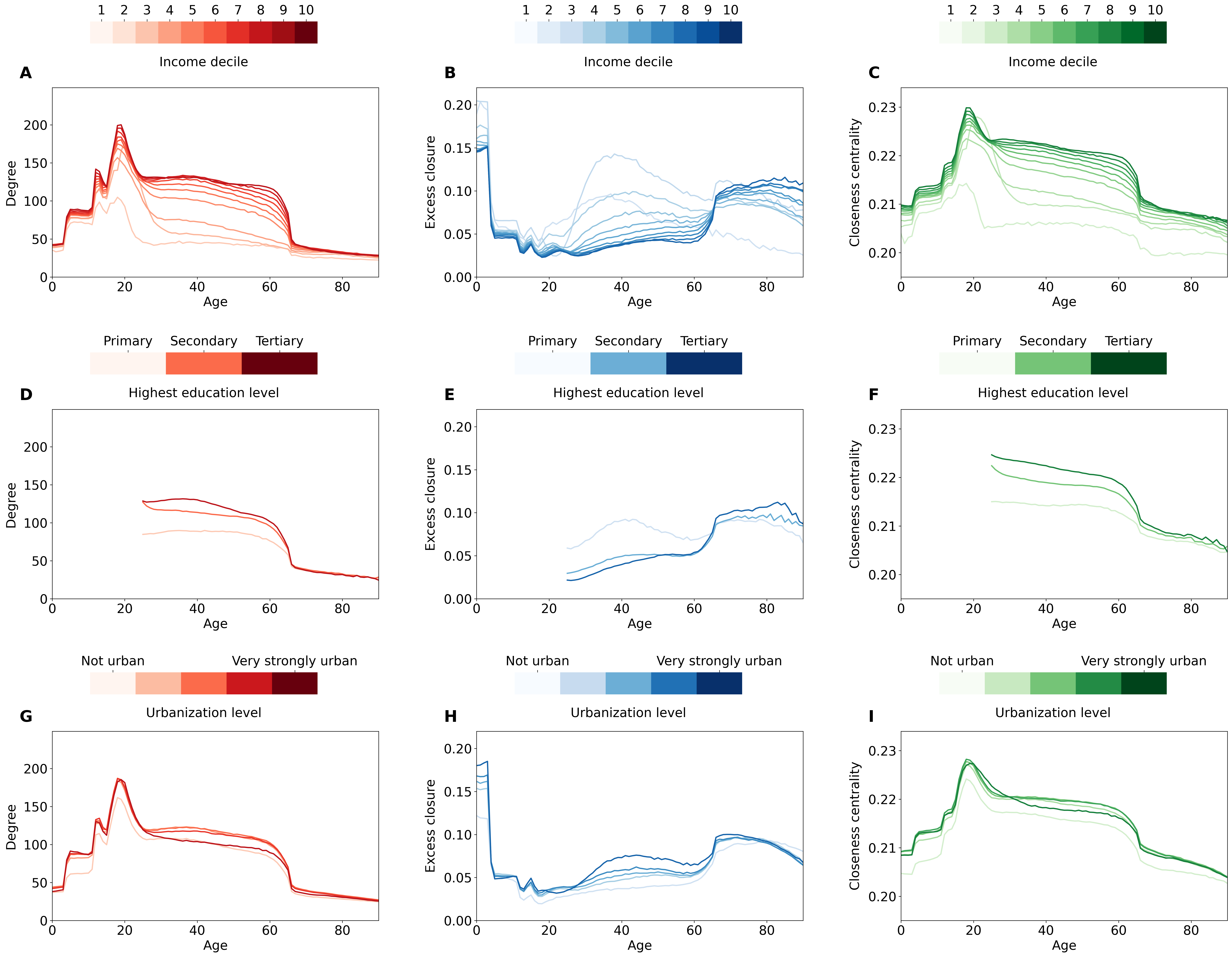}
    \caption{\textbf{Node properties for different demographic groups by age.} Average degree (red, first column), average excess closure (blue, second column), and average closeness centrality (green, third column) for people with different income (A-C), education levels (D-F), or type of living environment (G-I) as a function of age in the Dutch population-scale social network.}
    \label{fig:demography}
\end{figure}

Figure~\ref{fig:demography} shows how degree, excess closure, and closeness centrality (columns) of all the nodes change with age for different demographic groups by income, education, and urbanization level (rows) in the Dutch population. For young children and young adults, step-wise increases in degree are driven by their participation in different educational settings (Figure~\ref{fig:demography}A), with for example primary school and high school starting the ages of 4 and 12, respectively. After that, it is the size of the workplace that is dominant, with degree from work colleagues disappearing for retired generations. Throughout all ages, higher household income means higher degree, with differences in degree being particularly large between the first few income deciles. The mid-life difference in degree reflects a connection between income and work, i.e., people not working have neither income nor colleagues. Moreover, it indicates  that workplace size might be connected to financial compensation, with larger companies or organizations offering higher-paid jobs. Similar trends are observable for degree differences between education levels (Figure~\ref{fig:demography}D) and level of urbanization (Figure~\ref{fig:demography}G), that are both correlated with income. Degree patterns align well with the survey results from the meta-analysis of \cite{wrzus2013social}. Even though degree measures only the local structure in the direct neighborhood of a node, it seems to be correlated to the global reachability of all other network parts as measured by closeness centrality, which is defined as the inverse of the average distance to all other nodes (see \nameref{sec:methods}). The shape and variation across the different groups with respect to closeness are similar to that of the degree. 

In parallel to having a high degree, young children have high excess closure (Figure~\ref{fig:demography}B) since they are only part of family, neighborhood, and household structures. Subsequent phases of education and working opportunities come with decreasing excess closure which reaches its minimum around the university age. Working years are characterized by a slight increase in closure for higher-income groups, and a steep increase for the lowest income deciles. This is followed by increased closure in retirement years. High closure for low-income groups in adult life might be an important indicator signalling less access to new opportunities or information based on their immediate network structure \parencite{ aral2011diversitybandwidth}. Breakdown by education level and urbanization level for excess closure also follows the patterns shown by the income deciles, though it seems that there are much smaller differences in people's social opportunity structures by their living environment than by their socio-economic status \parencite{hofferth1998social}. Patterns for excess closure remain the same after controlling for degree, indicating that the observations are not only consequences of the $c(k)$ curve of Figure~\ref{fig:closure}B.

In summary, we find that degree, excess closure, and closeness centrality vary across demographic groups as well as for different age groups. These variations are important factors in understanding the role of social networks in individual socio-economic failure and success. The social embeddedness of economic action has been a cornerstone of socio-economic theory at least since the pioneering work of Granovetter \parencite{granovetter1985economic}. The rich empirical life-course network data now allows us to universally test these assumptions beyond survey-based research approaches and unravel how social structures in which individuals are embedded help or hamper their socio-economic development. 

\section{Discussion}

We mapped and analysed the social opportunity structure of an entire population based on high quality register data. This novel approach overcomes the persistent problem of selection bias in large-scale social network studies and is an efficient encompassing approach for mapping social network structures. We illustrated how the anatomy -- the overall structure -- of the population-scale social network  measures meaningful opportunity structures throughout people's lives. Specifically, we revisited three of the most commonly studied structural network measures in the field, being degree, distance, and closure, and presented markedly different results than expected based on prior work. 

Even though the degree distributions of the different layers resemble the fat-tailed degree distributions common in large-scale online social networks, these disappear once we consider the total degree distribution. The total degree distribution shows that nodes have both a lower and an upper bound on the number of connections. Almost none of the nodes have a degree of zero, and 97\% of the nodes have a degree of at least 20. This reflects  the social reality that it is very hard for people not to have social ties. This stands in contrast to degree distributions in most  large-scale communication networks or online social networks that measure an abundance of low-degree nodes with only 1 or 2 connections \parencite{mislove2007measurement}, which arguably reflects perceived node activity rather than actual node connectivity.

In our approach, we precisely map formal ties but do not have information on informal social ties  such as friendships or sport club memberships. One could argue that omitting friendship ties in a society where kinship is less and less important \parencite{david-barrett2019network} creates a biased view of the actual social structures. However, a large portion of the strong ties that people have come from the specific layers that we can capture with our data (see, e.g., \cite{vaneijk2010unequal} or \cite{buijs2022friends} for Dutch ego networks, and \cite{wrzus2013social} for a meta-analysis on survey-based personal network research). Moreover, multiple strong friendship ties actually would come from a longitudinal aspect of formal ties --- oftentimes, old high-school friendships or colleagues are one's current close friends. Thus, it would be an interesting future line of research to look at longitudinal population-scale social networks that would cover these strong ties.

We regard our network as contact \emph{opportunity structures} of people: formal ties that an individual can use as resources at some point of their career or personal lives. As in any large-scale social network study, it is difficult to assess how active ties are, e.g., whether two people connected actually meet or talk on a regular basis, or whether ties can actually provide advice, support or information. In the close family, extended family, and household layers, people either meet each other daily, or have a kinship connection that exists regardless of whether they maintain a very active, or a more distant relationship with each other. We suppose that some of the next-door relationships are active, and that in small workplaces, people tend to know all of their colleagues. Even if large university classes or the large workplace ties do not capture a realized relationship, in most cases, it is likely that even in a large workplace, one knows a colleague through another colleague. Therefore, we can interpret large degree values as an upper bound for the actual number of social ties. They are thus a proxy of how many formal network neighbors the nodes are able to reach in either 1 or 2 steps through their existing direct relationships. It also means that in the low-degree regime of Figure~\ref{fig:degree_distribution}C, the degree distribution coincides with what we would observe in actually realized formal social relationships. 
In this network of formal ties, there is a typical range of numbers of connections that people tend to have. This is somewhat in line with the 150 connections for people predicted by Dunbar \parencite{dunbar1992neocortex}. The lack of high-degree large hubs has implications when modelling dynamical phenomena such as epidemic spreading over contact opportunity networks.

Our new measure of \textit{excess closure} offers a solution for capturing parts of the multilayer population-scale network with a particularly strong sense of community. It can be interpreted as the extent of overlap and local bridging that is encoded into the multilayer structure of an individual's social opportunity structure. Higher values for excess closure thus indicate a stronger sense of actual community in ego networks. It also resolves the problem that the local clustering coefficient increases with increasing degree. Therefore, excess closure is a useful measurement of social closure structures in the wider opportunity structure, and in a generic manner applicable whenever it is possible to classify a node's connections into multiple different contexts (i.e., layers). The study of the anatomy of our population-scale social network also revealed that strong network wormhole edges as suggested by \cite{park2018strength} are conspicuously absent  in our network. Rather, co-affiliations in school classes and small workplaces wire together the fabric of family and household connections and as such create shortcuts between previously distant network parts. 

To conclude, we showed that it is possible to study the social opportunity study of an entire population based on register data, which presents a novel and highly efficient way of mapping meaningful patterns of social interaction. We revisited how degree, shortest paths, and  closure look like in this very particular multilayer network structure and showed that contrary to most large-scale social networks, both low-degree and high-degree nodes are rare. We measured how different edge types contribute to the reduction of average shortest path lengths between nodes. We found that not the individual long-range edges (wormhole edges), but rather co-affiliations in schools and small workplaces drive shrinking distances. These in return lead to a small-world structure, despite lack of large hubs or network wormhole edges. Because local clustering coefficients in this network tend to be high, we proposed a new measure called {excess closure}, that captures the overlap and local bridging between edges of different types in individuals' ego networks. 
The degree distributions, observed shortest path lengths, and excess closure all imply that it is crucial to distinguish between edges from different contexts (i.e., layers) in social networks, as this knowledge gives fundamental insights into the social opportunity structure of a whole population. We have shown how degree and closure evolve throughout an individual's life, and how it differs between demographic groups, and suggested a link between measurable network opportunity structures and socio-economic status. Our work offers insights in the interpretation of representative and population-scale networks derived from administrative register data, and shows that a network science viewpoint is useful in assessing social opportunity structures of different societal groups. Future comparative work across countries will be required to unravel the intricacies of the relation between social network structure and socio-economic opportunities. We therefore hope our work inspires studies of administrative register data in other countries that have similarly detailed yet hitherto unexplored population-scale social network.

\section{Data and Methods}
\label{sec:methods}

In this section we discuss the construction of the network nodes and layers and introduce the network measures used throughout the \nameref{sec:results} section.

\subsection{Nodes}

The data used in this paper is derived from multiple different registers of Statistics Netherlands (CBS) in which nodes are all 17.2M people registered in the Netherlands (in the Basisregistratie Personen, BRP) on October 1, 2018. 

\subsection{Layers}

The available layers are Close Family $C$, Extended Family $E$, Household $H$, Work $W$, School $S$, and Neighbors $N$. To avoid possible confusion of next-door neighbor relationships with network neighbors, i.e., two nodes connected by an edge, we are going to refer to relationships from the Neighbor layer as Next-door neighbors, which preserves the layer abbreviation $N$. 

\paragraph{Close family.}

The close family layer denoted by $C$ has been created from the complete digital parent-child register of the Netherlands and partner links.
The latter are derived from multiple different sources such as marriage registers or tax declarations, and include married partners, fiscal partners and co-parents having a common child. For details on the databases, and on the addition of sibling links, we refer the reader to \parencite{vanderlaan2022person}.
Only nodes and corresponding edges that are actively registered on 1st October 2018 are left in the data. 
Using the complete register for the derivation of family links means that even if the parents are deceased, sibling relationships can be observed.
Parent-child edges are directed, but if A is a parent of B, then B is a child of A, therefore, including both relationships creates an undirected close family layer. Sibling and partner relationships are undirected by definition.

\paragraph{Extended family.}
The extended family layer denoted by $E$ includes further family relationships that are deterministically added using parent-child relations and partner links. 
Extended family includes grandparent/grandchild, aunt-uncle/niece-nephew, first cousin, mother-father-in-law/son-daughter-in-law, sister-brother-in-law, stepparent/stepchild, and step-sibling relationships. 
All edges in this layer are undirected.

\paragraph{Household.} A household is a group of people living and managing their lives together at the same address. Using the assignment of people to households generated from different registers (BRP, partnerships, and addresses), housemate links connect all people assigned to the same household.
The data includes the category of institutional households, for example boarding houses, rescue homes, jails, orphanages, or care homes. These are of course larger in size than regular households.

\paragraph{Work.} 
The work layer contains links between colleagues that work for the same employer.
This is based on a process where every person is assigned an employer of their major source of income.
This means that each person is assigned at most one employer. After that, for companies that have less than 100 employees, every colleague link between these employees is created. 
However, if companies are larger than that, colleagues are sampled so that each person is linked the geographically closest 100 colleagues. We consider all created links to be undirected.

\paragraph{School.} The school layer includes information on where somebody goes to school, and aggregates information from different agencies on education. 
The data distinguishes between five levels of education in the Netherlands: elementary, secondary, secondary special, vocational, and higher.
We include connections between people in the same school, year, location, and type of education. In addition, the particular study programme is used in the case of universities. 

\paragraph{Next-door neighbors.} Next-door neighbors are relationships based on the household addresses of people. For each person, the 10 closest households are assigned as neighbors within 50~meter distance of their own address. If there are more than 10 such households, then they are randomly sampled down to 10. We assume all next-door neighbor edges to be undirected.

\subsection{Network measures}

We introduce the abstract description of our network dataset following the terminology set out by \textcite{kivela2014multilayer}.
We represent the data as an undirected multilayer graph $G = (V,E,L)$, where $V$ is the set of $n = |V|$ nodes, in our case, people of the Netherlands ($n=17.2\mathrm{M}$). The set of undirected edges $E$ connects pairs of nodes according to the above described relationships: \[E \subseteq \left\{\left(\left\{u,v\right\},\ell\right) : u,v\in V, u \neq v, \ell\in L\right\},\] 

There are six layers $L = \{C,E,H,S,W,N\}$, and for the number of links $|E|$ we use $m$. 
Our multilayer graph is \emph{node-aligned}: nodes are identical across all layers, and there are exclusively intra-layer edges. We can represent the edges with the \emph{reduced adjacency tensor} $A_{uv\ell}$, that is a generalization of the adjacency matrix, and encodes edges in the following manner:
\begin{equation}
    A_{uv\ell}=\left\{\begin{array}{ll}
    1 & \mbox{if }u\mbox{ and }v\mbox{ are connected in layer }\ell\in L,\\
    0 & \mbox{otherwise.}
\end{array}\right.
\end{equation}

Because the network is undirected, for all $u, v \in V$ it holds that $A_{uv\ell} = A_{vu\ell}$. 
We can also think about our network as an edge-coloured multigraph, that allows multiple types of relationships between two nodes. This representation can be viewed as a "flattened" version of the multilayer representation. 

\paragraph{Degree.}

The degree $k_{u,\ell}$ of a node $u \in V$ in a given layer $\ell\in L$ is given by
\begin{equation}
    k_{u,\ell} = \sum_{v \in V} A_{uv\ell}. 
\end{equation}

The total degree across all layers is 
\begin{equation}
    k_u = \sum_{\ell\in L} k_{u,\ell} = \sum_{\ell\in L} \sum_{v \in V} A_{uv\ell},
\end{equation}
which counts edges that are present in more layers multiple times.

The number of unique neighbors connected to node $u$ is
\begin{equation}
    k'_u = \left|\left\{v\in V: \sum_{\ell\in L}A_{uv\ell} >0 \right\}\right|.
\end{equation}

In data used to construct the family layer, there are missing parent-child records. Parent-child links exist in the dataset only if both people have been registered at some point in the databases after 1st October 1995. This means that especially for older people, we miss (derived) family relations. If a person’s parents were not  alive in 1995, we are unable to retrieve the siblings, nieces, nephews etc., because that parent-child link has not been recorded in the register. But it also means that those younger relatives do not have their aunts/uncles or cousins listed in the network. In addition, for first generation migrants, parents are not registered in the Netherlands and therefore not in our data. However, it is a realistic assumption that they have less kins in the country they immigrated into.

With the above considerations, we create a weighing scheme that accounts for the missing family links that affect 16.6\% of the nodes. For first-generation migrants, native people and second-generation migrants born after the 1995 register cutoff, we set the weights to 1 (36.4\% of the nodes). For all other people (46.9\% of the nodes), we create age groups of 10 years, and set the weight to the fraction of people with complete parent information in the group and in the population. We create averages and distributions for degrees using these weights. However, we cannot extend the same inference method for embeddedness, tie range, or closure, that we measure based on the links given in the data.

\paragraph{Embeddedness and tie range}
 
We measure embeddedness by counting the number of triangles an edge is part of \parencite{shi2007networks, sridharan2011statistical}. For an edge between nodes $u$ and $v$, this is formally given by
\[embeddedness_{uv} = \sum_{\ell_1,\ell_2\in L} \sum_{w \in V} A_{uw\ell_1}\cdot A_{wv\ell_2}.\]

Following \textcite{park2018strength}, the tie range of an edge between nodes $u$ and $v$ is the length of the second shortest path between $u$ and $v$. The shortest path length would be 1, since there is a direct edge, thus, we can also reformulate this distance as the shortest path between nodes $u$ and $v$ if the edge would not exist between them. Tie range is also one less than the length of the shortest cycle going through the edge. 
Any edge having an embeddedness larger than zero has a tie range of 2, since they are part of at least one triangle. Also, edges for which either end node has a degree of 1 have a tie range of infinity.

We measure tie ranges for edges that have zero embeddedness, and for which both end nodes have a degree larger than zero. We alternate two Breadth-First-Search (BFS) algorithms starting from the two endpoints with the other endpoint excluded from the search. 
On the first occasion when the frontier node sets of the two BFS's intersect, we stop the algorithm, and sum the actual search depths of the two trees.
 
\paragraph{Components and shortest path lengths.}

A path connecting two nodes in a multilayer network can consist of intra-layer edges that connect nodes of different layers, and inter-layer edges that connect nodes within a layer. Because our network is node-aligned, intra-layer edges are trivial edges, and we do not count these towards path lengths. 

A connected component of a graph is a subgraph, in which any two nodes are connected by a path. A connected component might consist of a single node, called an isolated node, in case the node's degree is 0, i.e., it is not connected to any other node in the network. 
The connected component with the largest number of nodes is called the giant component. 
 
We use the \texttt{teexgraph} library \parencite{takes2011determining} to compute the measures in Table~\ref{tab:sp}, including the exact diameter, being the longest shortest path length, as well as the average shortest path length, based on a sample of $1000 \cdot n$ node pairs in the network.

\paragraph{Closeness centrality.}

Closeness centrality is defined as the inverse of the average distance of a node $u$ to all other nodes:

\begin{equation}
    closeness\_centrality_u = \frac{n-1}{\sum_{v \in V, u\neq v} d_{uv}},
\end{equation}

Here, $d_{uv}$ is the length of the shortest path between nodes $u$ and $v$. The shortest path is calculated using all edges from all layers.
We employ a fast approximation for a 0.03\% node sample of the largest connected component~\parencite{eppstein2006fast}.

\paragraph{Local clustering coefficient.}

The the local clustering coefficient of a node $u$ is given by:
\[
c\_local_u = \frac{\left|\left\{ (u,v,w) : \left\{u,v\right\},\left\{v,w\right\},\left\{u,w\right\}\in E'\right\}\right|}{k'_u\cdot(k'_u-1)},
\]
which is the number of closed triangles over the number of neighbor pairs around node $u$, if $E'=\left\{\left\{u,v\right\}: \sum_{\ell\in L} A_{uv\ell}>0\right\}$ (essentially omitting edge layer identifiers for readability), meaning that there is at least one edge between nodes $u$ and $v$.

\paragraph{Excess closure.}

Just measuring the local clustering coefficient would capture mostly the share of intra-layer triangles, since we expect edges from the same layer to form clustered structures around the egos. Kinship networks up to the second degree are the merges of extremely highly clustered structures through partnerships. Household, school and work relationships are complete networks because they come from bipartite projections, and next-door person relationships are clustered because of spatial proximity. Triangles are considered from the viewpoint of the ego in an ego network, and triangle types are characterized by a triplet $\{\ell_1,\ell_2,\ell_3\}$, where $\ell_1$ and $\ell_3$ correspond to layers of edges to two selected neighbors of the ego (ordered by ascending node labels), and $\ell_2$ corresponds to the layer type of the third, closing edge of the triangle. 

\begin{figure}[ht!]
\centering
\includegraphics[width=0.8\textwidth]{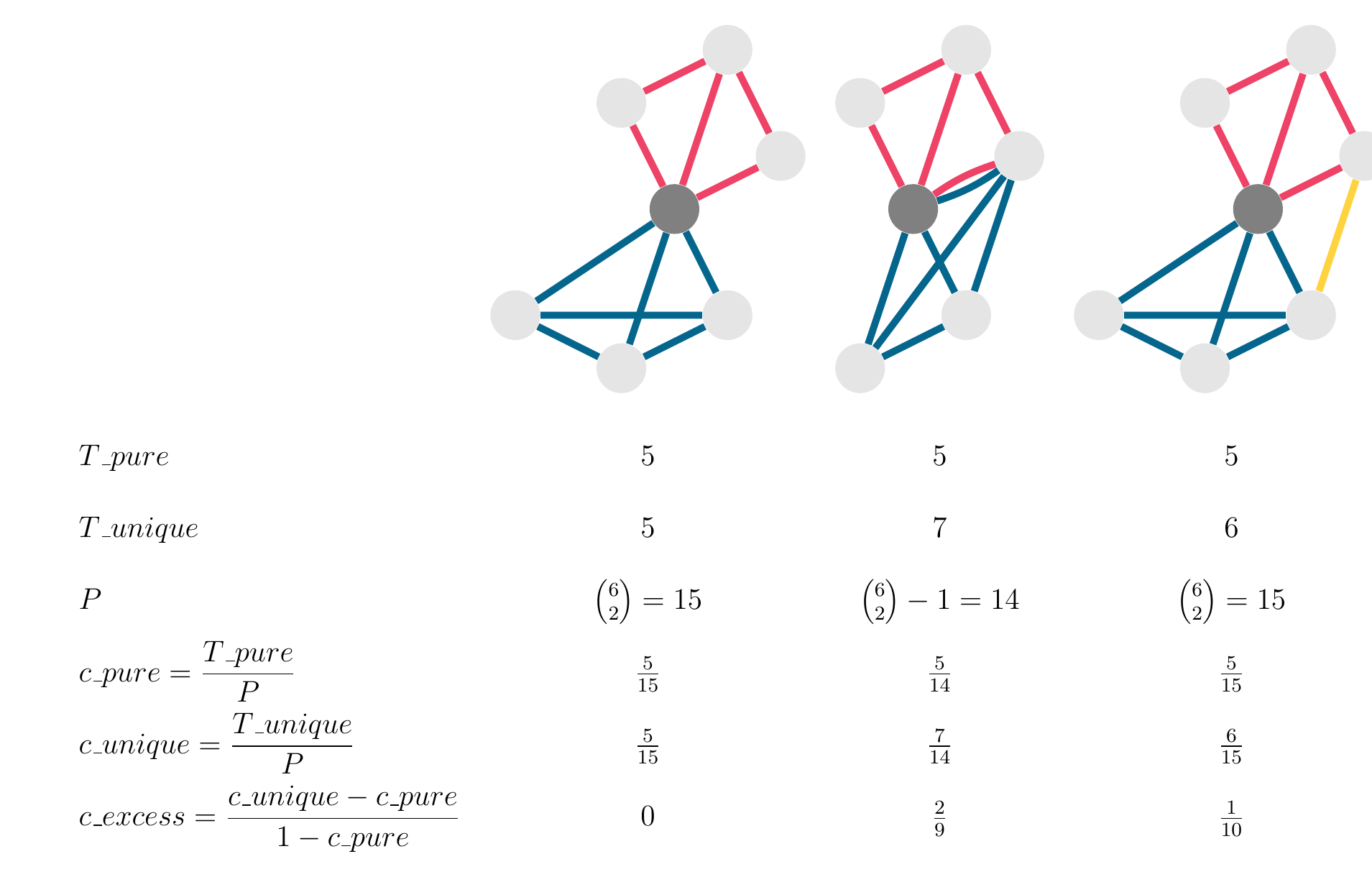}
\caption{\textbf{Three sample ego networks illustrating the concept and the calculation of excess closure.} Ego is the central node shaded darker, edges come from three layers color-coded by red, blue, and yellow. The values of $T\_{pure}$, $T\_{unique}$, $c\_{actual}$, $c\_{pure}$, $P$, and $c\_{excess}$ are in the table below the figures. }
\label{fig:excess_closure_example}
\end{figure}

For this, we first sum up the number of triangles $T_{\ell\ell\ell}$ around an ego for which all three edges come from the same layer $\ell$: $$T\_{pure} = \sum_{\ell\in L} T_{\ell\ell\ell}.$$
Note that this definition might count some node triplets twice if they exist in multiple layers.
Next, we divide this sum by the number of possible different alter tie pairs around the ego $P$. This is different from the number of unique neighbor pairs, since as in Figure~\ref{fig:excess_closure_example}b, an alter can be connected to the ego in multiple layers, thus, the same neighbor pair can be formed using only intra-layer edges, or also edges from different layers. But $P$ also accounts for the fact that a neighbor pair cannot be formed from two edges that lead to the same alter.

Formally, for a node $u$, the number of alter tie pairs $P_u$ can be calculated as
\[
P_u = {k_u \choose 2} - \sum_{v \in {neighbors_u}} {\sum_{\ell\in L} A_{uv\ell} \choose 2},
\] 
where $k_u$ is the total degree of node $u$, and $neighbors_u = \left\{v \in V : \sum_{\ell\in L} A_{uv\ell}>0\right\}$.

We define the pure clustering coefficient $c\_{pure} =  T\_{pure} / P$, which would actually be a theoretical minimum clustering coefficient, if there would be no multi-edges between the ego and any of its neighbors, and there would not be any cross-edges between neighbors that are linked by edges from different layers. To put it differently, it is the case when the ego’s neighbors of different layers are never overlapping and never linked to each other.

Then, we calculate the number of unique closed triangles $T\_{unique}$ around the ego by listing each node triplet only once, even if there are multiple possible layer combinations that close wedges between the same nodes. This would include to some extent the previously calculated $T\_{pure}$ triangle count, and in addition, all other triangles of mixed layer types. We divide this number also by the number of possible neighbor pairs $P$, thus calculating the clustering coefficient $c\_{actual} =  T\_{unique} / P$. We obtain the excess closure $c\_{excess}$ by comparing where this clustering is between $c\_{pure}$ and the theoretical maximum clustering of 1:

\begin{equation}
    c\_{excess} = \frac{c\_{actual} - c\_{pure}}{1 - c\_{pure}}.
\end{equation}

Figure~\ref{fig:excess_closure_example} shows the concept for three example egos with edges from three different layer types. The table below the figure enumerates the number of unique closed triangles $T\_{unique}$, $T\_{pure}$, $P$, and then $c\_{actual}$, $c\_{pure}$, and $c\_{excess}$.

From the viewpoint of the ego, excess closure relates to the number of closed triangles that consist of edges from different layers. However,  we still cannot be sure whether this measure captures overlap or cohesion in the ego’s different  opportunity structures, or it actually captures bridges between two very different social circles, such as a family member and a colleague being neighbors of each other. Having a multilayered social network, we can actually characterize the triangle types from this angle. Each triangle can contribute to cohesion, and/or to bridging. Cohesion happens when the triplet $\{\ell_1,\ell_2,\ell_3\}$ encoding the triangle has $\ell_1=\ell_3$ from the very same layers. For example, in the case when the ego’s two family members are also household members of each other, triangle $\{F,H,F\}$, the edge from layer $H$ does not bridge multiple social contexts of the ego. Bridging is characterized by cases when $\ell_1\neq \ell_2$ and $\ell_2\neq \ell_3$, for example, in the triangle $\{S,N,W\}$, where the ego’s school tie and work tie are neighbors of each other. But in some sense, this type of closure also causes the ego’s social circles to “overlap”, therefore, it still counts towards excess closure, too.

The measure of \emph{excess closure} introduced for capturing clustering in the opportunity structures belonging to the different network layers is related to multiple other clustering coefficient definitions from the literature. \textcite{barrett2012taking} draw attention to the fact that the multilayer nature of triadic closure has to be taken into account when calculating clustering. Partial clustering coefficients are introduced in \textcite{baxter2016cycles} to capture the fraction of triangles with edges from 1, 2, or 3 layers. The paper of \textcite{cozzo2015structure} proposes so-called layer-decomposed local clustering coefficients, based how many layers edges of multilayer three-step walks come from. In their notation, our excess closure can be written as\[c\_{excess} = \frac{c^{(2)}+c^{(3)}}{1-c^{(1)}},\] where $c^{(i)}$ corresponds to the fraction of walks consisting of $i$ different types of inter-layer edges.

\section*{Ethical statement}

The register dataset used for creating the Dutch population-scale network is pseudonymized before research access, and it does not contain personal information such as address or date of birth. All data storage and analysis was done within the secure server environments of Statistics Netherlands. Further details about ensuring data security and privacy are described in \textcite{vanderlaan2022whole}.

\section*{Acknowledgements}

We are thankful for the collaboration with Statistics Netherlands (CBS) colleagues Jan van der Laan, Edwin de Jonge, and Gert Buiten. The project has been funded by Platform Digitale Infrastructuur Social Sciences and Humanities (\url{http://www.pdi-ssh.nl}). We would also like to thank the POPNET team (\url{https://www.popnet.io}), in particular Yuliia Kazmina, for helpful suggestions and discussions. Moreover, we are grateful for insightful comments from Márton Karsai and Matteo Magnani on early versions of the manuscript.

\subsection*{Conflicts of interest and data and code availability}

The authors declare no conflicts of interest. The whole population network file for 2018 is included in the Statistics Netherlands Microdata Catalogue as a beta product and is available for access through the Statistics Netherlands Microdata Services Remote Access Environment (RA); see \parencite{vanderlaan2022person}. The Python library used as a basis for the analysis in this paper is available via GitHub: \url{https://github.com/popnet-io/popnet_mln}.

\subsection*{Author contributions}

E.B., E.H., and F.T. conceptualized the study, developed the methodology, and wrote the manuscript. E.H. and F.T. acquired funding, supervised and administered the project. E.B. and F.T. developed the software. E.B. performed the analysis and the investigations. E.B. visualized the results. All authors reviewed and approved the manuscript.

\renewcommand*{\bibfont}{\small}

{
\singlespacing
\printbibliography
}
\end{document}